\documentclass{elsart}

\usepackage{amssymb}
\usepackage{amsmath}
\usepackage{graphicx}
\newcommand{\be}{\begin{equation}}
\newcommand{\ee}{\end{equation}}
\newcommand{\ben}{\begin{eqnarray}}
\newcommand{\een}{\end{eqnarray}}
\newcommand{\lb}{\label}

\begin{document}

\begin{frontmatter}

\title{Finite-size effects on the chiral phase diagram of
four-fermion models in four dimensions}

\author[UFRB]{L.M. Abreu\corauthref{cor1}}
\ead{lmabreu@ufrb.edu.br}
\author[CBPF]{A.P.C. Malbouisson}
\ead{adolfo@cbpf.br}
\author[UFBA]{J.M.C. Malbouisson},
\ead{jmalboui@ufba.br}
\author[UNB]{A.E. Santana}
\ead{asantana@fis.unb.br}

\corauth[cor1]{Corresponding author}

\address[UFRB]{Centro de Ci\^{e}ncias Exatas e Tecnol\'{o}gicas,
Universidade Federal do Rec\^{o}ncavo da Bahia, 44380-000, Cruz das
Almas, BA, Brazil}
\address[CBPF]{Centro Brasileiro de Pesquisas F\'{\i}sicas/MCT,
22290-180, Rio de Janeiro, RJ, Brazil}
\address[UFBA]{Instituto de F\'{\i}sica, Universidade Federal da
Bahia, 40210-340, Salvador, BA, Brazil}
\address[UNB]{Instituto de F\'{\i}sica, Universidade de
Bras\'{\i}lia, 70910-900, Bras\'{\i}lia, DF, Brazil}

\begin{abstract}

We study the size dependence of the dynamical symmetry breaking in
the four-dimensional Nambu-Jona-Lasinio model. We show that the
presence of boundaries reduces the chiral breaking region, and this
effect is strengthened for a larger number of compactified
dimensions. A critical value for the length of the compactified
dimensions exists, below which the dynamical symmetry breaking is
not possible. Considering finite temperature and chemical potential,
the chiral phase structure for the system with compactified
dimensions is obtained. A gradual decreasing of the chiral breaking
region with increasing of chemical potential is found. Also, at
fixed chemical potential, the decreasing of the size of the system
changes the order of the chiral phase transition.

\end{abstract}

\end{frontmatter}

\section{Introduction}

The phase structure of the strongly interacting matter attracts a
great deal of interest. Due to the intrincate mathematical structure
of Quantum Cromodynamics (QCD), effective models that incorporate
some of its properties have been largely employed. In this sense,
four-fermion models, as the Nambu-Jona-Lasinio (NJL)
model~\cite{NJL}, are very useful for the investigation of dynamical
symmetries when the system is under certain conditions, like finite
temperature, finite chemical potential, gravitational field, etc.
\cite{Kl,HK,Bu}. In particular, an interesting aspect in the
analysis of the phase transitions of four-fermion models is the
effect of space
compactification~\cite{Kim,HCGL,KS,BS,GG,KKK,AGS,EKTZ}. Boundary
effects have also been considered for quark-meson
models~\cite{BKP,BKPR}. The general question is to estimate the
relevance of the fluctuations due to finite-size effects in the
phase diagram. With this purpose, different approaches have been
used to study various aspects of these models, as the finite-size
scaling analysis~\cite{KS}, the multiple reflection
expansion~\cite{KKK}, and the zeta-function method~\cite{AGS}.

In this paper, we extend the techniques introduced in
Ref.~\cite{AGS} and investigate finite-size effects on the dynamical
symmetry breaking of the four-dimensional NJL model at finite
temperature and chemical potential. This is done in the framework of
zeta-function regularization and compactification methods~\cite{EE}.
This approach allows in a simple way to determine analytically the
size-dependence of the effective potential and the gap equation.
Then, phase diagrams at finite temperature and chemical potential,
where the symmetric and broken phases are separated by
size-dependent critical lines, are obtained. Note that our approach leads to exact
results within the mean-field (Hartree) theory without making use of
any additional approximation. This allows to avoid, for example, difficulties
appearing in calculating the density of states using the multiple reflection expansion,
as pointed out in Ref.~\cite{KH}.

We organize the paper as follows. In Section~II, we calculate the
effective potential of the NJL model in the mean-field
approximation, using the zeta-function method. The size-dependent
gap equation is discussed in Section~III, while the phase diagrams
are shown and analyzed in Section~IV. Finally, Section~V presents
some concluding remarks.

\section{The formalism}

Our starting point is the massless version of the NJL model,
described by the Lagrangian density, 
\be \mathcal{L} = \bar{q}  i\!\!
\not{\!\partial}  q + \frac{G}{2} \sum ^{N^2-1}_{a=0} \left[ \left(
\bar{q} \alpha ^a q \right)^2 + \left( \bar{q} i \gamma _5 \alpha ^a
q \right)^2 \right], 
\lb{NJL} 
\ee 
where $q$ and $\bar{q}$ are the
$N$-component spinors, and the matrices $\alpha ^a$ are the
generators of the group $U(N)$, with $\alpha ^0 =
\mathbf{I}/\sqrt{N}$.

We perform the bosonization assuming that only one auxiliary field,
that associated with the bilinear $\bar{q} \alpha ^0 q $, takes
non-vanishing values. This auxiliary field, denoted here by
$\sigma$, plays the role of a dynamical fermion mass, such that when
it has a non-vanishing value, the system is in the chiral broken
phase. We shall work in the Euclidian space, performing the Wick
rotation in the time coordinate. For generality, we consider the
$D$-dimensional Euclidean space-time, restricting latter to the
$D=4$ case. Besides, taking $\sigma$ uniform, i.e. independent of
coordinates, we obtain the effective potential up to one-loop order,
at leading order in $\frac{1}{N}$, as 
\be \frac{1}{N} U_{eff}  =
\frac{ \mathcal{A}_{eff}}{V} = \frac{ \sigma ^2}{2G} + U_{1} (\sigma),
\lb{effpot} \ee where $\mathcal{A}_{eff}$ is the effective action, $V$
is the volume and \be U_{1} (\sigma) = - h_D \int \frac{d^D
k}{(2\pi)^D} \ln{\left( \; \frac{k_E^2 + \sigma^2}{\lambda ^2}
\right)}. 
\lb{oneloop} 
\ee 
Above, $h_D$ is the dimension of the
Dirac representation and $\lambda$ is a scale parameter.

To take into account finite-size effects on the phase structure of
this model, the system is considered to have $d$ ($\leq D$)
compactified dimensions. We denote the Euclidian coordinate vectors
by $x_E$ and write $x_E = (y,z)$, where
\be
y=\left( y_1 \equiv x_E^1, ..., y_n \equiv x_E ^n \right), \;\; z=\left( z_{1} \equiv x_E
^{n+1}, ..., z_{d} \equiv x_E ^{D}  \right),
\ee
with the $z_j$
component being defined in the interval $[0, L_j]$ and $n=D-d$,
corresponding to the topology
$\mathbb{R}^n\times\mathbb{S}^{1_{n+1}}\times
\cdots\times\mathbb{S}^{1_{D}}$ where $\mathbb{S}^{1_j}$ is a
circumference of radius $L_j$. Therefore, the compactification of
the $z$-coordinates makes the $k_z$-components of the momentum $k_E$
to assume discrete values,
\be
 k_{z}^j \rightarrow
\frac{2\pi}{L_j}(n_{j} + c_j ), 
\lb{prescription}
\ee
where $ n_j = 0, 1, 2, \ldots$ and $c_j = \frac{1}{2} \; (j=1,2,...,d)$ for
antiperiodic boundary conditions. Note that we may associate a given
$L_{j}$ with the inverse of temperature $\beta = 1/T$, say $L_{j_0}
\equiv \beta$, thus treating the system at finite temperature with
$d-1$ compactified spatial dimensions. In this case, a finite
chemical potential, $\mu$, can also be introduced through the rule
$c_{j_0}=\frac{1}{2} - \frac{i\beta \mu}{2 \pi}$.

In the following, we use the zeta-function regularization
method~\cite{EE}, rewriting Eq.~(\ref{oneloop}) as  
\be 
U_{1}(\sigma) = \frac{h_D}{2V} \left[ \zeta ' (0) + \ln{\lambda ^2}\zeta
(0)\right], 
\lb{U1} 
\ee 
where $\zeta  (s)$ is given by \be \zeta (s)
=  V_n \sum_{n_{1},...,n_{d} =  -\infty}^{+\infty } \int \frac{d^n
k_y}{(2\pi)^{n}}\left[k_z ^2 + k_y ^2  + \sigma^{2} \right]^{-s },
\lb{zeta1} \ee where $k_z ^2 = \sum_{j=1}^{d}
\frac{4\pi^2}{L_{j}^{2}} \left( n_j + c_j \right)^2$, with $V_n$
being the $n$-dimensional volume. The techniques of dimensional
regularization can be used to perform the integration over the
non-compactified $k_y$-components, yielding
 \be 
 \zeta \left(s;\left\{a_j\right\}, \left\{c_j\right\} \right) =
\frac{V_n}{(4\pi)^{n/2}} \frac{\Gamma \left( s- \frac{n}{2}
\right)}{\Gamma \left( s \right) } Y_{d}^{\sigma^2}\left(
s-\frac{n}{2};\left\{a_j\right\}, \left\{c_j\right\} \right),
\lb{zeta2} 
\ee 
where $Y_{d}^{\sigma^2} \left( \nu ;
\left\{a_j\right\}, \left\{c_j\right\} \right)$ is the generalized
Epstein-zeta function, defined by \be Y_{d}^{\sigma^2}\left( \nu ;
\left\{a_j\right\}, \left\{c_j\right\} \right) =
\sum_{n_{1},...,n_{d} = -\infty} ^{+\infty }\left[ a_{1} \left(n_{1}
+ c_1 \right)^{2}+\cdots +a_{d} \left(n_{d} + c_d
\right)^{2}+\sigma^{2} \right]^{-\nu }, \lb{epstein1} \ee with $a_j
= 4 \pi ^2/L_j ^2$. Note that $Y_{d}^{\sigma^2}$ is well-defined
only for $\rm{Re} \; \nu > d/2$, but it can be analytically
continued, becoming a meromorphic function, into the whole complex
$\nu$-plane.

As remarked in Refs.~\cite{EE,AGS}, the analysis of the pole
structure of the zeta-function implies that Eq.~(\ref{U1}) must be
written as 
\be 
U_{1} \left(\sigma ;\left\{a_j\right\},
\left\{c_j\right\} \right) =  \frac{h_D}{2V_d (4\pi)^{n/2} } \Gamma
\left( - \frac{n}{2} \right) Y_{d}^{\sigma^2}\left(
-\frac{n}{2};\left\{a_j\right\}, \left\{c_j\right\} \right),
\lb{U1odd} 
\ee 
for $n$ odd, or 
\ben 
U_{1}\left(\sigma ;\left\{a_j\right\}, \left\{c_j\right\} \right) & = &
\frac{h_D}{2V_d (4\pi)^{n/2} }
\frac{(-1)^{\frac{n}{2}}}{\frac{n}{2}!} \left\{ Y_{d}^{\sigma^2 \;'}
\left( -\frac{n}{2};\left\{a_j\right\}, \left\{c_j\right\} \right)
\right. \nonumber \\
&+ & \left. Y_{d}^{\sigma^2}\left( -\frac{n}{2};\left\{a_j\right\},
\left\{c_j\right\} \right)\left[ \ln{\lambda^2} - \gamma - \psi
\left(\frac{n}{2} + 1 \right) \right] \right\}, 
\lb{U1even} 
\een 
for $n$ even, where $\gamma$ and $\psi(s)$
denote the Euler-Mascheroni constant ($\approx 0.5772$) and the
digamma function, respectively. Both expressions will be used in the
discussion of symmetry breaking in the four-dimensional NJL model
with compactified dimensions.

\section{The gap equation}

The phase structure of the model is studied through the gap
equation, obtained by minimizing the effective potential with
respect to $\sigma$, \be \left.\frac{\partial  }{\partial
\sigma}U_{eff} \left(\sigma ;\left\{a_j\right\}, \left\{c_j\right\}
\right)\right|_{\sigma = m} = 0, \lb{gap} \ee where $m$ is the order
parameter of the chiral phase transition, that is the dynamically
generated fermion mass. We now discuss the size effects on the gap
equation starting, for completeness and to set the free space
parameters, by treating the model in absence of boundaries.

\subsection{Non-compactified model}

Let us consider the NJL model in four dimensions without
compactified dimensions. This corresponds to $d=0$, and so $n=D=4$.
This system has been treated in the literature (see
Refs.~\cite{IKM,Z} for reviews) using other methods. With the
zeta-function approach, the renormalized effective potential is
given by \be \frac{1}{N} U_{eff} \left(\sigma; d=0 \right) =
\frac{\sigma ^2}{2G_R}+ \frac{h_D (D-1)}{ (4\pi)^{D/2} } \Gamma
\left( 1 - \frac{D}{2} \right) \lambda^{D-2} \sigma ^2 -
\frac{h_D}{D (4\pi)^{D/2} } \Gamma \left( 1 - \frac{D}{2} \right)
\sigma^D , \lb{effren} \ee where the renormalized coupling constant
$G_R$ is defined by \be \frac{1}{G_R} = \frac{1}{G} - \frac{h_D
(D-1)}{ (4\pi)^{D/2} } \Gamma \left( 1 - \frac{D}{2} \right)
\lambda^{D-2}. \lb{GR1} \ee  Note that Eq.~(\ref{effren}) is valid
for $ 2 \leq D < 4$; it is singular for $D=4$, due to the pole of
the gamma-function. In the limit $D\rightarrow 4$, we define
$\epsilon = 4 - D $ and obtain an analytically regularized effective
potential, taking $h_D=4$, as
 \ben \frac{1}{N} \frac{U_{eff} \left(\sigma \right)}{
\lambda ^4} & = & \frac{1}{2 G_R} \frac{\sigma ^2}{ \lambda ^4} -
\frac{6 \sigma ^2}{(4 \pi)^2 \lambda^2}\left( \frac{1}{\epsilon} -
\gamma + \ln{4 \pi} + \frac{1}{3} \right) \nonumber  \\
& + &  \frac{\sigma ^4}{(4 \pi)^2 \lambda^4} \left(
\frac{1}{\epsilon} -\gamma + \ln{4 \pi} + \frac{3}{2} -
\ln{\frac{\sigma ^2}{\lambda^2}}\right); \lb{U1d0} \een this shows
explicitly the singular behavior of Eq.~(\ref{effren}) as $\epsilon
\rightarrow 0$.

Let us compare Eq.~(\ref{U1d0}) with the corresponding expression
obtained by using the cut-off regularization~\cite{IKM,Z}, \be
\frac{1}{N} \frac{U_{eff} \left(\sigma  \right)}{ \lambda  ^4} =
\frac{1}{2 G} \frac{\sigma ^2}{ \lambda  ^4} - \frac{6 \sigma ^2}{(4
\pi)^2 \lambda ^2}\left( \ln{\frac{\Lambda ^2}{\lambda ^2}}-
\frac{2}{3} \right) + \frac{\sigma ^4}{(4 \pi)^2 \lambda ^4}
\left(\ln{\frac{\Lambda ^2}{\lambda ^2}} + \frac{1}{2} -
\ln{\frac{\sigma ^2}{\lambda ^2}}\right), \lb{U1d0cutoff} \ee where
$\Lambda $ is the cut-off parameter which must be larger than
$\lambda $. Thus, the zeta-function and cut-off methods are
equivalent through the correspondence \be \frac{1}{\epsilon} -\gamma
+ \ln{4 \pi} + 1 \leftrightarrow \ln{\frac{\Lambda ^2}{\lambda^2}}
\lb{correspondence}. \ee Hence, with the use of the correspondence
in Eq.~(\ref{correspondence}), the non-trivial solution of the gap
equation derived from Eq.~(\ref{U1d0}) can be written as
 \be \frac{1}{G_c} - \frac{1}{G_0} = - \frac{1}{ m \lambda  ^{2}}
 \left. \frac{\partial  }{\partial \sigma}U_{1}
 \left(\sigma \right)\right|_{\sigma = m} =
 \frac{4 m ^2}{(4 \pi)^2 \lambda ^2} \ln{\frac{\Lambda ^2}{m ^2}}, \lb{gap1}
\ee where we have defined the dimensionless coupling constant $G_c =
\lambda^{2} G_R$, and \be
 \frac{1}{G_0} = \left.\frac{\partial  }{\partial \sigma}U_{eff}
 \left( \sigma \right)\right|_{\sigma \rightarrow 0} =
 \frac{12}{(4 \pi)^2} \left( \ln{\frac{\Lambda ^2}{\lambda^2}} -
 \frac{2}{3} \right). \lb{G0}
\ee In Eq.~(\ref{gap1}), it is possible to identify the constant
$G_0$ acting as a critical value; when $ G_c > G_0 $ we have
dynamically generated fermion mass. A value for $G_0$ can be fixed
by choosing values for the mass scale $\lambda$ and the cut-off
$\Lambda$ from phenomenological arguments; note that, for $D=4$, the
NJL model has to be considered as an effective model.

\subsection{Presence of boundaries}

To take into account temperature and finite-size effects, we must
analyze the modified gap equation, 
\be 
\frac{1}{G_c} - \frac{1}{G_0}= -  \frac{1}{ \overline{m} \lambda ^{2}} \left.\frac{\partial
}{\partial \sigma}U_{1} \left(\sigma ;\left\{a_j\right\},
\left\{c_j\right\} \right)\right|_{\sigma = \overline{m}}, 
\lb{gap2}
\ee 
where $\overline{m} = \overline{m}(\{ a_j \},\{ c_j \})$ is the
boundary modified fermion mass. Then, accordingly to the
zeta-function approach, using Eqs.~(\ref{U1odd}) and (\ref{U1even}),
the modified gap equation, Eq.~(\ref{gap2}), is written as 
\be
\frac{1}{G_c} = \frac{1}{G_0} + \frac{4}{ \lambda
^{2}V_d(4\pi)^{n/2} } \Gamma \left( 1 - \frac{n}{2} \right)
Y_{d}^{\overline{m}^2 }\left( -\frac{n}{2}+1; \left\{a_j\right\},
\left\{c_j\right\} \right), \;\; \rm{ for \;\textit{d}=1,3 };
\lb{gap3} 
\ee 
while it has the form 
\ben 
\frac{1}{G_c} & = &
\frac{1}{G_0} + \frac{4}{ \lambda ^{2} V_2 4\pi } \left\{
Y_{2}^{\overline{m}^2 \;'}  \left( 0;\left\{a_j\right\},
\left\{c_j\right\} \right) \right. \nonumber \\
&+ & \left. Y_{2}^{\overline{m} ^2}\left( 0 ;\left\{a_j\right\},
\left\{c_j\right\} \right)\left[ \ln{\lambda ^2} - \gamma - \psi
\left( 1 \right) \right] \right\}, \;\; \rm{ for \;\textit{d}=2 };
\lb{gap4} 
\een 
and finally, 
\be 
\frac{1}{G_c} = \frac{1}{G_0} -
\frac{4}{ \lambda ^{2} V_4 } \, {\rm FP}  \left[
Y_{4}^{\overline{m}^2 }\left( 1; \left\{a_j\right\},
\left\{c_j\right\} \right) \right], \;\; \rm{ for \;\textit{d}=4},
\lb{gap5}
 \ee 
 where ${\rm FP}[Y_{4}^{\overline{m}^2}]$ means the
finite part of $Y_{4}^{\overline{m}^2 }$.

Thus, taking the fermion mass approaching to zero in
Eqs.~(\ref{gap3}), (\ref{gap4}) and (\ref{gap5}), we obtain the
critical values of the coupling constant $G_c$ with the corrections
due to the presence of boundaries, for the cases $d=1,3$, $d=2$ and
$d=4$, respectively. In this context, $\left. Y_{d}^{\overline{m}^2}
\right|_{\overline{m}^2 \rightarrow 0}$ reduces to a homogeneous
generalized Epstein zeta-function $Y_{d}$.

The next step is the construction of analytical continuations for
$Y_{d}$, which are written through a generalized recurrence
formula~\cite{AGS}, \ben Y_{d} \left( \nu; \left\{a_j\right\},
\left\{c_j\right\} \right) & = & \frac{\Gamma\left( \nu -
\frac{1}{2}
 \right)}{ \Gamma(\nu)}  \sqrt{\frac{\pi}{a_d}} Y_{d-1}
 \left(\nu - \frac{1}{2};\left\{a_{j \neq d}\right\},
 \left\{c_{j \neq d}\right\} \right) \nonumber \\
 & +&  \frac{4 \pi^s }{\Gamma(\nu)} W_d \left( \nu - \frac{1}{2};
 \left\{a_j\right\}, \left\{c_j\right\} \right),
\label{epstein4}
\een
where the symbol $\left\{a_{j \neq d}\right\}$ means that the
parameter $a_d$ is excluded from the set $\left\{a_{j}\right\}$, and
\begin{equation}
W_d \left( \eta ; \left\{a_j\right\}, \left\{c_j\right\} \right) =
\frac{1}{\sqrt{a_d} } \sum_{\{n_{j \neq d} \in \mathbb{Z}\}}
\sum_{n_d=1}^{\infty}\cos{\left(2\pi n_d c_d \right)} \left( \frac{
n_d}{\sqrt{a_d} X_{d-1}}  \right)^{\eta} K_{\eta}\left( \frac{2\pi
n_d}{\sqrt{a_d} } X_{d-1} \right) \; ; \label{Wd}
\end{equation}
in the above equation
$X_{d-1}=\sqrt{\sum_{k=1}^{d-1}a_{k}\left(n_{k} + c_k \right)^{2}} $
and $K_{\nu} (z)$ is the modified Bessel function of second kind
(see Refs.~\cite{Kirsten,Abreu} for a discussion of the $a_{j}
\leftrightarrow a_{l}$ symmetry).

\section{Phase structure}

\subsection{Finite-size effects on the critical coupling}

We now analyze the $L_j$-dependent critical curves for the phase
diagram from the gap equation in the limit $\overline{m} \rightarrow
0$. First, we study the compactification of spatial coordinates at
zero temperature. Thus, we take Eq.~(\ref{gap3}) in the cases $d=1$
and $d=3$, and Eq.~(\ref{gap4}) for the $d=2$ case. After that, we
use the recurrence formula, given by Eq.~(\ref{epstein4}), and
perform the necessary manipulations.

Considering the simplest situation in which all spatial coordinates
are restricted to intervals of the same length and obey antiperiodic
boundary conditions, i.e. $L_j=L$ and $c_j =1/2$ for $j=1,\dots,d$,
we obtain the following gap equations \be \frac{1}{G_c} =
\frac{1}{G_0} - \frac{A_d }{ (L \lambda)^2}, \lb{Gc} \ee where \ben
A_1 & =&  \frac{ 1}{6} \approx 0.16, \lb{A1} \\
A_2 & =&  A_1 - \frac{4}{\pi} \sum_{n_1 = - \infty}^{\infty}
\sum_{n_2 = 1}^{\infty}(-1)^{n_2} \left( \frac{\left| n_1 +
\frac{1}{2}\right| }{n_2} \right) ^{\frac{1}{2}}
K_{-\frac{1}{2}}\left( 2 \pi n_2 \left| n_1 + \frac{1}{2}\right|
\right)
\approx 0.22,\nonumber \\ \lb{A2}   \\
 A_3 & =&  A_2 - \frac{4}{\pi} \sum_{n_1,n_2 = - \infty}^{\infty}
 \sum_{n_3 = 1}^{\infty}(-1)^{n_3}  K_{0}\left( 2 \pi n_3
 \sqrt{\sum_{i=1}^{2}\left( n_i +
 \frac{1}{2}\right)^2  }\right)
 \approx 0.26.
 \lb{A3}
 \een

\begin{figure}[th]\begin{center}
\includegraphics[{height=6.0cm,width=8.0cm}]{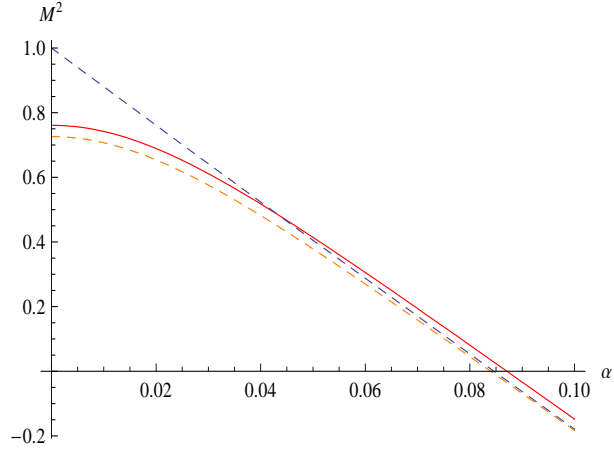}
\caption{The critical coupling constant $G_c$, Eq.~(\ref{Gc}), as a
function of $x = (L \lambda )^{-1}$.  The dashed, dotted and solid
lines correspond to the cases of $d=1,2,3$ compactified dimensions,
respectively, with $A_d$ given by Eqs.~(\ref{A1}--\ref{A3}). We
fixed $G_0 \simeq 5.66$.}\end{center} \label{FIG1}
\end{figure}

To illustrate the results above, in Fig.~\ref{FIG1} we plot the
critical coupling constant $G_c$, given by Eq.~(\ref{Gc}), as a
function of $x = (L \lambda )^{-1}$, with the coefficient $A_d$
taking the values given by Eqs.~(\ref{A1}), (\ref{A2}) and
(\ref{A3}); we also fix the value of $G_0 \simeq 5.66$, obtained
from Eq.~(\ref{G0}) by choosing $\Lambda \approx 1.25 \,{\rm GeV}$
and $\lambda \approx 280 \,{\rm MeV}$~\cite{VR}. The three cases
considered correspond, respectively, to the system between two
parallel planes a distance $L$ apart; in the form of an infinite
cylinder having a square transverse section of area $L^2$; in the
form of a cubic box of volume $L^3$. For each situation the chiral
breaking region corresponds to the region above the corresponding
line. Since $G_c$ increases as $L$ decreases, we can infer that the
finite-size effects require a stronger interaction to maintain the
system in the chiral breaking region, as $L$ is diminished.

%%%%%%%%%%%%%%%%%%%%%%%%%%%%%
%%%%%%%%%%%%%%%%%%%%%%%%%%%%%

We also find from Eq.~(\ref{Gc}) that, for each $d$, a critical
value $L_c$ exists below which the chiral breaking region is
completely suppressed; this critical value is obtained by setting
the right hand side of Eq.~(\ref{Gc}) equal to zero. For $G_0 \simeq
5.66$, considering the values of $A_d$ given by
Eqs.~(\ref{A1}--\ref{A3}), we get $x _c= (L_c \lambda )^{-1}\simeq
1.03$ for $d=1$, $x _c\simeq 0.89$  for $d=2$ and $x _c\simeq 0.83$
for $d=3$. We see that the cubic box has the greatest value of
$L_c$, which means that the largest number of compactified
dimensions is the scenario which needs the strongest interaction to
keep the system in the chiral breaking region.

It is worth noting that the behavior of the critical coupling constant as a function
of $L$ does not depend on the number of compactified dimensions, as it is expected
from finite-size scaling arguments.

\subsection{System at finite temperature}

Now we consider the system with compactified spatial dimensions and
at finite-temperature. Since the way of introducing finite-size
effects is through a generalized Matsubara prescription, we can
identify one of the compactified dimensions with the Euclidian
(imaginary) time and take the compactification length as the inverse
of the temperature, say $L_1 = \beta \equiv T^{-1}$. So, to analyze
the $(L,T)$-dependent phase diagram, we must look at the critical
equation obtained from Eqs.~(\ref{gap1}) and (\ref{gap2}), \be
\frac{4 m^2}{ (4 \pi)^2 \lambda ^2} \ln{\frac{\Lambda ^2}{m^2}} +
\frac{1}{ \overline{m} \lambda^2 } \left.\frac{\partial  }{\partial
\sigma}U_{1} \left(\sigma ;\left\{a_j\right\}, \left\{c_j\right\}
\right)\right|_{\sigma = \overline{m}} =0, \lb{critical} \ee where
$a_1=4 \pi^2 / L_1^2 \equiv 4 \pi^2 / \beta ^2$ and, for simplicity,
we take $a_2=\cdots=a_d=4 \pi^2 / L^2$. Then, we look at
Eq.~(\ref{gap4}) for $d=2$, Eq.~(\ref{gap3}) for $d=3$, and
Eq.~(\ref{gap5}) for $d=4$ considering, in all these cases, $L_1 =
\beta$. Next, we perform the necessary manipulations, in an
analogous way as we have done in the previous subsection, and take
the limit $\overline{m}\rightarrow 0$. Thus, using
Eqs.~(\ref{epstein4}) and (\ref{Wd}), we obtain \be \frac{4 m^2}{ (4
\pi)^2 } \ln{\frac{\Lambda ^2}{m^2}} - \frac{A_1}{L^2 } +
\frac{4}{\pi L } \sum_{n_1 = - \infty}^{\infty} \sum_{n_2 =
1}^{\infty}(-1)^{n_2} \left( \frac{\left| n_1 + \frac{1}{2}\right|
}{n_2 \beta L  } \right)^{\frac{1}{2}} K_{-\frac{1}{2}}\left( 2 \pi
n_2 \frac{\beta}{ L}\left| n_1 + \frac{1}{2}\right| \right)  = 0,
\lb{ft2} \ee for $d=2$; \ben & & \frac{4 m^2}{ (4 \pi)^2 }
\ln{\frac{\Lambda ^2}{m^2}} -
\frac{A_2}{L^2 } \nonumber \\
& & + \frac{4}{\pi L^2} \sum_{n_1, n_2 = - \infty}^{\infty}
\sum_{n_3 = 1}^{\infty}(-1)^{n_3}  K_{0}\left( 2 \pi n_3
\frac{\beta}{ L}\sqrt{ \left( n_1 + \frac{1}{2} \right)^2 + \left(
n_2 + \frac{1}{2} \right)^2 } \right) = 0, \nonumber \\
\lb{ft3} 
\een
for $d=3$; and finally 
 \ben & & \frac{4 m^2}{ (4 \pi)^2 } \ln{\frac{\Lambda
^2}{m^2}} - \frac{A_3}{L^2 } + \frac{4}{\pi L^3} \sum_{n_1, n_2, n_3 = - \infty}^{\infty}
\sum_{n_4 = 1}^{\infty}(-1)^{n_4}
 \left( \frac{ n_4 \beta L }{\sqrt{\sum_{i=1}^{3} \left( n_j +
 \frac{1}{2} \right)^2 }} \right)^{\frac{1}{2}} \nonumber \\
& & \times K_{\frac{1}{2}}\left( 2 \pi n_4 \frac{\beta}{ L}\sqrt{\sum_{i=1}^{3}
 \left( n_j + \frac{1}{2} \right)^2} \right)= 0, 
\lb{ft4} 
\een 
in the case of $d=4$, where $A_d$, for $d=1,2,3$, are
given by Eqs.~(\ref{A1})--(\ref{A3}).

In Fig.~\ref{FIG2}, we plot the phase diagrams corresponding to
Eqs.~(\ref{ft2}--\ref{ft4}) in the $(x,T)$-plane, where $x$ and $T$
are the inverse of the compactification length and the temperature,
respectively, measured in units of $m$, that is $x= (L m )^{-1}$ and
$ T = (\beta m )^{-1}$. Each critical line separates the chiral
breaking region, below the line, from the chiral restoration phase,
above the line.

%%%%%%%%%%%%%%%%%%%%%%%%%%%%%%
%\begin{figure}[th]\begin{center}
%\includegraphics[{height=8.0cm,width=8.0cm}]{fig2.eps}
%\caption{ The phase diagram in the $(x,T)$-plane, where $x = (L m
%)^{-1}$ and $ T = (\beta m )^{-1}$. Dashed, dotted and solid lines
%represent Eqs.~(\ref{ft2}), (\ref{ft3}) and (\ref{ft4}), for
%$d=2,3,4$ compactified dimensions, respectively. For each case the
%non-trivial mass phase corresponds to the region below the
%corresponding line. The ratio between the parameters $\Lambda $ and
%$m$ is taken as $\frac{\Lambda}{m} = 4.46$~\cite{VR}.}\end{center} \label{FIG2}
%\end{figure}
%%%%%%%%%%%%%%%%%%%%%%%%%%%%%

In the limit $ x \rightarrow 0$, corresponding to the system without
spatial boundaries, the critical temperature is $ T_c  \approx 0.68
\,m$. The size effects start to appear for $x \approx 0.3 $. We see
from the curves that the critical temperature decreases as the size
of the system diminishes. Our results indicates that there is a
minimal size of the system for the existence of a chiral broken
phase. The critical values of the compactification lengths for
suppression of the chiral breaking region are given by: $x _c= (L_c
m )^{-1}= 0.68$ for $d=2$, $x _c=0.58$ for $d=3$ and $x _c=0.54$ for
$d=4$. Note that, for the system between two parallel planes a
distance $L$ apart (the case $d=2$), the critical line is symmetric
by the change $ x \leftrightarrow T $, as expected. It can also be
seen that the critical temperature has a faster decreasing with the
size reduction for the system in the form of a cubic box. This fact
is confirmed by the greatest value of the critical length (the
length for which the critical temperature vanishes) in this case.

In order to obtain a quantitative estimate for our results, we
consider $m$ as the constituent quark mass, $m = 280$~GeV according
to Ref.~\cite{VR}. In this scenario, with the appropriate
conversions, the critical temperature for the system without spatial
boundaries is $ T_c \simeq 0.68 \, m \approx 190$~MeV and the size
effects begin to be noted for $L = (x m )^{-1} \approx 2.4 $~fm. For
this choice of $m$, the critical values of the compactification
length are $L_c = (x_c m )^{-1} \approx 1.1$~fm for $d=2$, $L_c =
1.23$~fm for $d=3$ and $L_c = 1.32$~fm for $d=4$.

\subsection{System at finite temperature and at finite chemical
potential}

The phase diagram can also be studied by taking into account the
dependence on the chemical potential. As previously remarked, this
is done by taking in the prescription (\ref{prescription}), for the
momentum associated with the compactified Euclidian time, $L_1
\equiv \beta = 1/T$ and $c_1=\frac{1}{2} - \frac{i\beta \mu}{2
\pi}$. Thus, considering the simplest situation, $d=1$, which
corresponds to the system in bulk (absence of boundaries), the use
of Eqs.~(\ref{gap1}) and (\ref{gap3}) leads to the following
equation for the critical line, \be \frac{4 m^2}{ (4 \pi)^2 }
\ln{\frac{\Lambda ^2}{m^2}} - \frac{1}{6 \beta^2 } - \frac{\mu^2}{2
\pi^2 } = 0. \lb{ft5} \ee Now, if we consider the system with
compactified spatial coordinates, the $\mu$-dependence appears as
additional factors in the last term of the left hand side of
Eqs.~(\ref{ft2}), (\ref{ft3}) and (\ref{ft4}), which are obtained
after the appropriate analytical continuation.

So far we have considered only second-order phase transitions.
However, the inclusion of a finite chemical potential can alter the
nature of the chiral phase transition. To understand as this happens
we proceed as follows. We expand the gap equation (\ref{gap2}) for
$\overline{m}$ near criticality, obtaining \be A(T,L_j,\mu) +
B(T,L_j,\mu) \overline{m}^2 + C(T,L_j,\mu) \overline{m}^4 = 0,
\lb{GL1} \ee where the coefficients $A(T,L_j,\mu)$, $B(T,L_j,\mu)$
and $ C(T,L_j,\mu)$ are defined according to
Eqs.~(\ref{gap3})--(\ref{gap5}), i.e. in terms of homogeneous
generalized Epstein zeta-functions $Y_{d}(\nu)$, with different
$\nu$-exponents. The tricritical point is then determined by
imposing the condition \be B(T,L_j,\mu) =0 . \lb{tp} \ee

\begin{figure}[th]\begin{center}
\includegraphics[{height=8.0cm,width=8.0cm}]{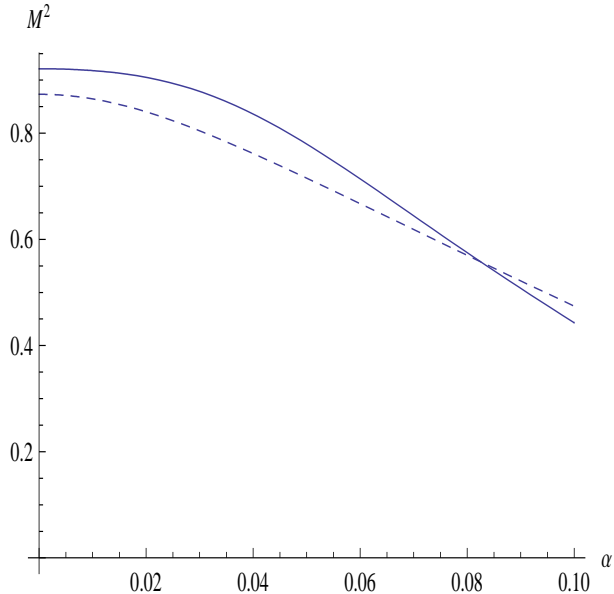}
\caption{ The phase diagram in the $(\mu,T)$-plane for the case
without boundaries ($d=1$), with $T$ and $\mu$ measured in units of
$m$. Solid and dashed lines represent the second-order and
first-order phase transition lines, respectively, and the dot is the
tricritical point. The chiral breaking region is below the line.
Here we take $\frac{\Lambda}{m} = 4.46$~\cite{VR}.} \end{center}\label{FIG3}
\end{figure}

As an example, in Fig.~\ref{FIG3} we plot the phase diagram in the
$(\mu,T)$-plane for the case without boundaries ($d=1$). We identify
the solid and dashed lines as representing, respectively, second- and
first-order phase-transition lines. The dot locates the tricritical
point, which occurs at $T_c \simeq  0.46 \, m \approx 128$~MeV and
$\mu_c \simeq 0.90 \, m \approx 252$~MeV. As expected, at high
temperature or low chemical potential, a second-order phase
transition is suggested, while in the low-temperature or
high-chemical potential limit the system experiments a first-order
transition.

%%%%%%%%%%%%%%%%%%%%%%%%%%%%%%
%%%%%%%%%%%%%%%%%%%%%%%%%%%%%

Finally, in Fig.~\ref{FIG4} we plot the phase diagram of the system
with three compactified spatial dimensions ($d=4$) in the
$(x,T)$-plane, for some values of the chemical potential. The chiral
breaking region is gradually diminished with increasing chemical
potential. So, when $\mu$ increases, lower values of temperature and
smaller sizes are necessary to reach a phase-transition line. In
addition, nonzero values of the chemical potential can alter the
order of the phase transition, as expected. Nevertheless, the
decreasing of the size can also changes the order of the phase
transition for the system at fixed $\mu$.

%%%%%%%%%%%%%%%%%%%%%%%%%%%%%%
%\begin{figure}[th]\begin{center}
%\includegraphics[{height=8.0cm,width=8.0cm}]{fig4.eps}
%\caption{The phase diagram in the $(x,T)$-plane, for some values of
%$\mu$ (in units of $m$), for the case of the system in a box
%($d=4$), where $x = (L m )^{-1}$ and $T = (\beta m )^{-1}$. Solid
%and dashed lines represent the second-order and first-order
%phase-transition lines, respectively, and the dot is the tricritical
%point. The chiral breaking region corresponds to the region below
%each line. Here we fixed $\frac{\Lambda}{m} \approx
%4.46$~\cite{VR}.} \end{center} \label{FIG4}
%\end{figure}
%%%%%%%%%%%%%%%%%%%%%%%%%%%%%

\section{Concluding Remarks}

In this work, we have studied dynamical symmetries of a four-fermion model
when the system is under certain conditions. We have investigated
finite-size effects on the chiral phase structure of the four
dimensional NJL model employing zeta-function regularization and
compactification methods. In the mean field approximation, this
allows to derive analytically expressions of the gap equation when
$d$ ($\leq 4$) dimensions are compactified.

First, the dependence of the critical coupling constant on the size
of the system was analyzed, comparing the cases of one, two and
three compactified spatial dimensions with the same compactification
length $L$. The presence of boundaries implies that a stronger
interaction is needed to maintain the system in the chiral broken
phase; this effect is stronger for greater number of compactified
dimensions. It is shown that there is a minimal size of the system
below which there is no chiral breaking region; the corresponding
critical values of the compactification lengths, for $d=1,2,3$, are
determined.

The chiral phase structure at finite temperature was investigated by
taking one of the compactification lengths, that associated with the
compactified Euclidian time, equal to the inverse of temperature.
The results suggest that, as the size of the system diminished,
finite-size effects start to appear for a given value of the
compactification length, with the temperature rapidly decreasing as
$1/L$ is further increased; this behavior is more accentuated for the
case with the greater number of compactified spatial dimensions,
giving the largest decreasing of the chiral breaking region.

The study is concluded with the analysis of the dependence of the
chiral phase transition on the chemical potential. For the bulk
system, i.e. the system without spatial boundaries, we get a simple
expression for the phase-transition line, showing that the critical
temperature decreases as the chemical potential increases. The
nature of the phase transition changes from second- to first-order
at specific values of $\mu$ and $T$ which locates the tricritical
point. For the system with compactified spatial dimensions, we find
that the chiral breaking region decreases as the chemical potential
is increased; that is, when $\mu$ increases, lower values of
temperature and smaller sizes are necessary to keep the system in
chiral broken phase. We show that, for a fixed value of $\mu$,
the decrease of the size leads to a change in the order of the
transition. 

Possible extensions of this work are the investigation of properties
of fermion-fermion condensates under finite-size conditions, as well
as, the study of the system under the influence of an external
magnetic field.

{\bf Acknowledgments}

This work received partial financial support from CNPq and FAPERJ,
Brazilian Agencies.

\end{document}